\documentclass{sig-alternate}
\usepackage{subfigure}
\usepackage{graphicx}

\makeatletter
\def\@copyrightspace{\relax}
\makeatother

\begin{document}

\title{Multipath-TCP in ns-3}

\numberofauthors{1}
\author{
\alignauthor
Morteza Kheirkhah, Ian Wakeman and George Parisis\\
\affaddr{Department of Informatics, University of Sussex, UK}\\
\email{\{m.kheirkhah, ianw, g.parisis\}@sussex.ac.uk}
}

\maketitle

\begin{abstract}
In this paper we present our work on designing and implementing an ns-3 model for MultiPath TCP (MPTCP). Our MPTCP model closely follows MPTCP specifications, as described in RFC 6824, and supports TCP NewReno loss recovery on a per subflow basis. Subflow management is based on MPTCP's kernel implementation. We briefly describe how we integrate our MPTCP model with ns-3 and present example simulation results to showcase its working state.

\end{abstract}

\section{Introduction}
MultiPath TCP (MPTCP) has been gaining significant attention during the past years. Its ability to spread network traffic to multiple subflows of a single MPTCP connection, in a way that fairness with other competing flows is preserved, promises more efficient usage of network resources and resilience in the face of network failures. In data centres, multi-homed servers, network path multiplicity and very high aggregated bandwidth are becoming the norm and transport protocols that could maximise the usage of network resources and minimise flows' completion time, while being fair with existing TCP flows, are heavily researched. Network simulations are the only realistic route to capture, examine and understand the behaviour of a transport protocol at the scale required in the context of data centres. Existing research is commonly based on custom-developed network simulators, instead of relying on and extending existing, well-tested, open-source code. This makes verifying the correctness of the simulation itself extremely difficult. Software bugs, even minor ones, can seriously affect the quality and reliability of results. In many cases, simulators' source code is not made available and, therefore, it is impossible to repeat and verify published results.

To date, there has been a single attempt to implement MPTCP in ns-3 version 3.6 \cite{MPTCP-NS3.6}. However, this model was never merged with any stable version of ns-3 and also became obsolete after TCP was rewritten in ns-3.8. In this model only a single client could connect to an MPTCP server; i.e. a server could not fork new MPTCP connections. This is a problem particularly when dealing with realistic traffic models in data centres. Additionally, the model did not support nodes running TCP and MPTCP in parallel, a feature that is often required in data centre experimentation, when evaluating fairness among competing TCP and MPTCP flows. MPTCP tokens \cite{RFC6824}, which uniquely identify MPTCP connections in a host and are used to associate new subflows to an existing MPTCP connection, were not used either. Finally, several other simplifications (e.g. the MPTCP connection and its subflows do not follow standard TCP state transitions) were present.

In this paper we present an ns-3 model of MPTCP, which is based on RFC 6824 \cite{RFC6824}. Subflow management is done similarly to the Linux kernel implementation and loss recovery is based on TCP NewReno. Our model overcomes all limitations described above with the ambition to become the official MPTCP model for ns-3. The source code will soon be submitted for review.

\begin{figure*}[!htb]
\subfigure[]{
\includegraphics[width=5.55cm, height=4.3cm]{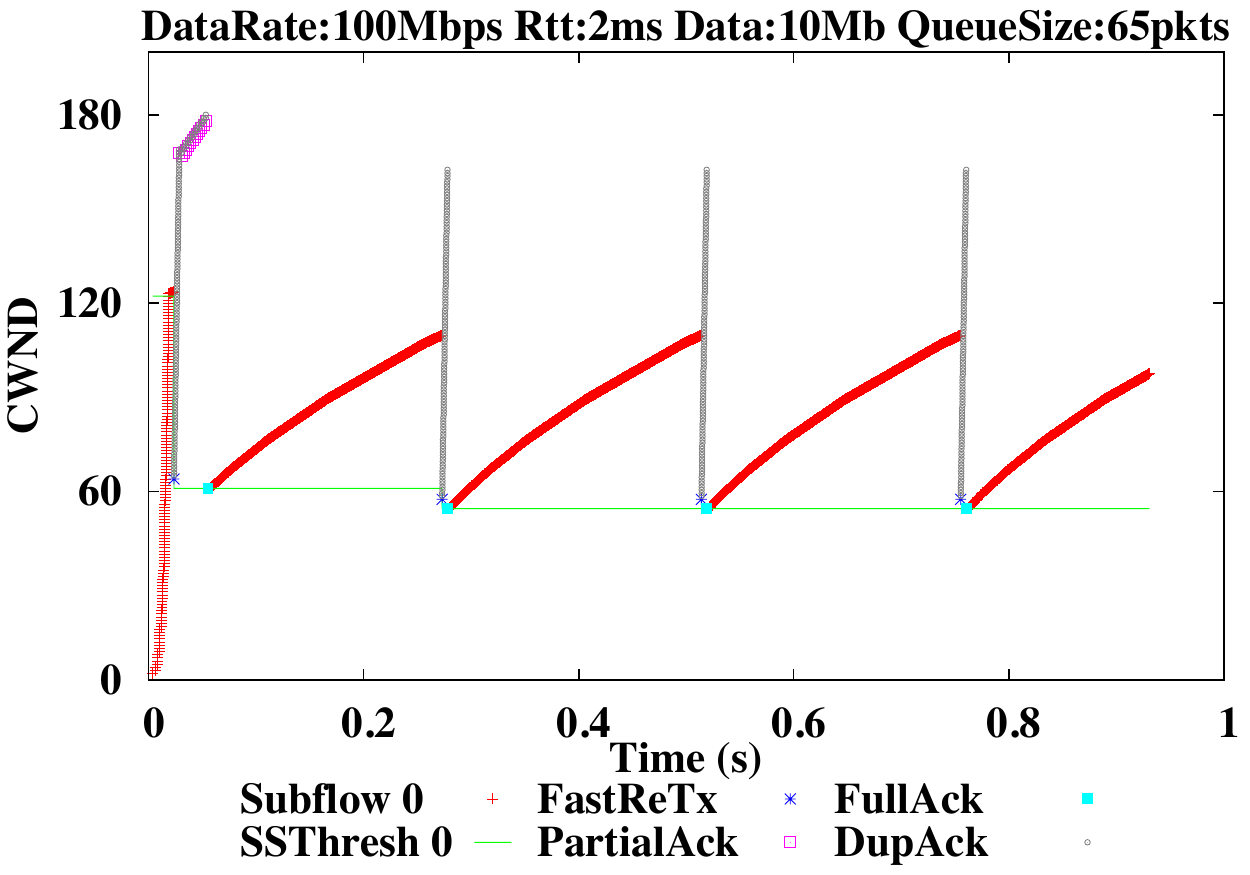}
\label{fig:sub1}}
\quad
\subfigure[]{
\includegraphics[width=5.55cm, height=4.3cm]{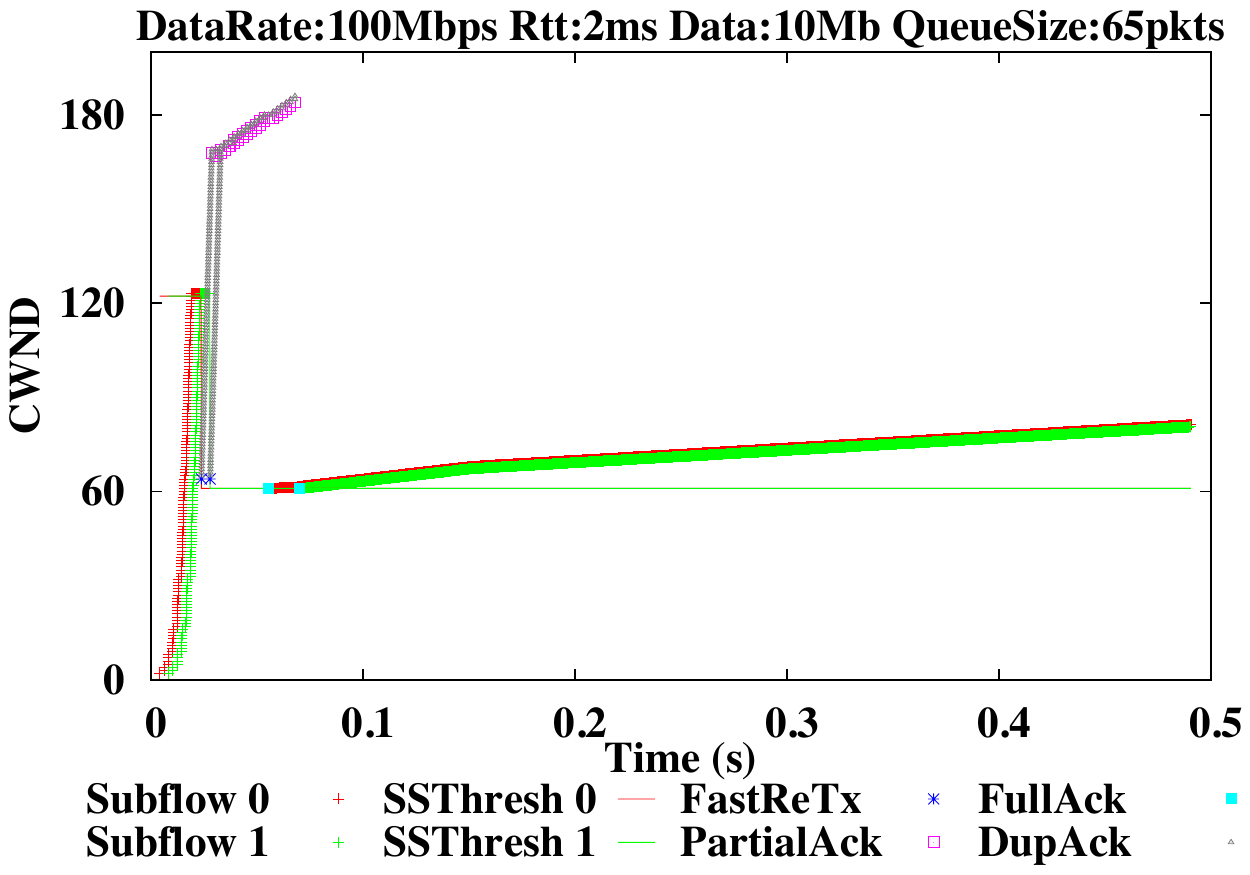}
\label{fig:sub2}}
\quad
\subfigure[]{
\includegraphics[width=5.55cm, height=4.3cm]{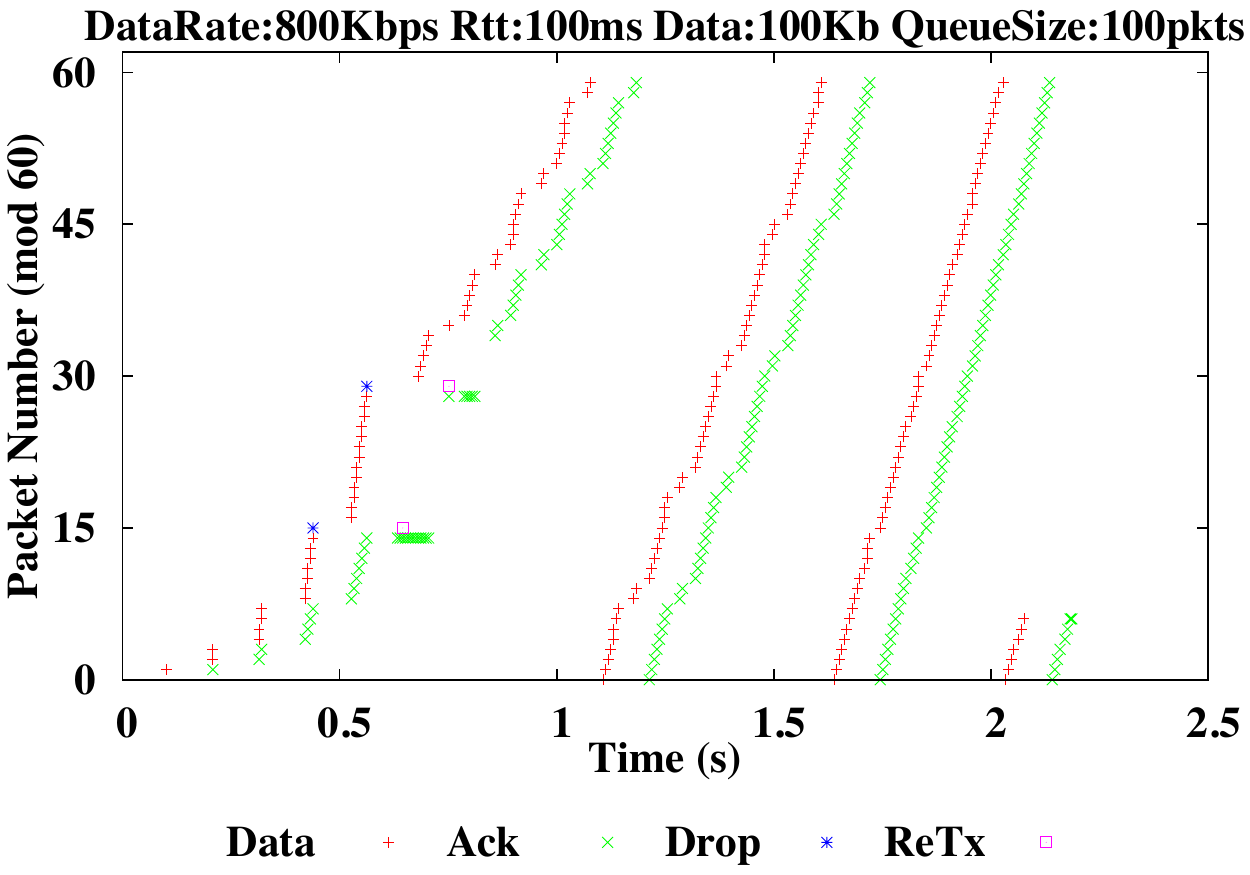}
\label{fig:sub3}}
\caption{(a) MPTCP with a single subflow, (b) MPTCP with two subflows (c) MPTCP  with a single subflow and two packet drops (as in \cite{Fall_comparisons}). MSS is 536 bytes.} 


\label{sdsdsd}
\end{figure*}

\section{MPTCP Model} \label{NewModel}
\subsection{Connection and SubFlow Management} \label{management}
Our MPTCP model follows the Linux kernel implementation of MPTCP \cite{MPTCP-Linux}. Each MPTCP connection can have several subflows, each of which operates as a regular TCP connection. Each connection starts with a master subflow, the only subflow presented to the application through the exported socket interface. In our model we have implemented the following classes:

\noindent\textbf{\textit{MpTcpSocketBase}.} This class implements the MPTCP control block and exports the socket API to ns-3 applications. It performs data scheduling, packet reordering, congestion control and loss recovery for all MPTCP subflows. On the server side, a MPTCP connection is represented with one listening \textit{MpTcpSocketBase} object which forks new \textit{MpTcpSocketBase} objects for each accepted MPTCP connection. On the client side an \textit{MpTcpSocketBase} object represents an MPTCP connection with a server. \textit{MpTcpSocketBase} is a subclass of \textit{TcpSocketBase}.

\noindent\textbf{\textit{MpTcpSubflow}.} This class represents an MPTCP subflow and is a subclass of the \textit{Object} class. An \textit{MpTcpSocketBase} object may have multiple \textit{MpTcpSubflow} objects.

\noindent\textbf{\textit{TcpL4Protocol}.} We have changed \textit{TcpL4Protocol} so that MPTCP connections can be handled, without disrupting any existing TCP functionality. As with single-path TCP models, this class is an interface between the transport and network layers and is responsible for sending and receiving packets to and from the network layer, respectively. When a packet is received, it looks up an \textit{Ipv4EndPoint} based on the TCP header's four-tuple. In our model, several \textit{Ipv4EndPoint} objects, representing endpoints for respective MPTCP subflows, can be associated to one MPTCP connection. MPTCP token support is implemented in this class as well. A token is a locally unique identifier assigned to an MPTCP connection upon establishment. When a sender initiates a new subflow, the receiver looks up an \textit{Ipv4EndPoint} based on the token passed in the \textsc{mp-join} option and forwards the request to the respective \textit{MpTcpSocketBase} object. Requests for new MPTCP connections are resolved using the four-tuple and forwarded to the listening \textit{MpTcpSocketBase} object. 

\subsection{MPTCP Signalling} \label{signalling}
Our MPTCP model closely follows the specifications set in RFC 6824\cite{RFC6824}. MPTCP signalling is implemented as follows: 

\noindent\textbf{Connection Establishment.} MPTCP connection establishment follows the standard TCP three-way handshake. The client attaches an \textsc{mp-capable} option in the \textsc{syn} packet to denote its MPTCP support. The \textsc{syn} packet is forwarded to the listening \textit{MpTcpSocketBase} object, which, in turn, forks a new \textit{MpTcpSocketBase} object to handle communication with the client. If the server supports MPTCP, an \textsc{mp-capable} option is attached to the \textsc{syn-ack} response. A randomly generated 32-bits token is carried in the \textsc{mp-capable} option. The token is mapped to the respective \textit{MpTcpSocketBase}; this mapping being stored in the \textit{TcpL4Protocol} object. 

\noindent\textbf{Subflow Establishment.} After connection establishment, communicating endpoints advertise available IP addresses to each other using the \textsc{add-addr} option. New subflows can be later established using a three-way handshake and attaching the \textsc{mp-join} option in the \textsc{syn} packet. The token is also attached so that the receiving side can resolve the subflow establishment request to an existing MPTCP connection, as described in the previous section.

\noindent\textbf{Sending and Receiving Data Packets.} MPTCP utilises two separate sequence number spaces, one per-connection (64-bits) and one per-subflow (32-bits). The former is used for packet reordering and loss recovery at connection level and it is signalled via the Data Sequence Signal (\textsc{dss}) option. The latter is used for the same reasons at a subflow level and is carried in the sequence number field of the TCP header.

\noindent\textbf{Connection Teardown.} As with standard TCP, each subflow terminates with a four way \textsc{fin} handshake. An MPTCP connection is also terminated at a connection level by signalling a \textsc{data-fin} option. In our model, connection teardown is performed at a subflow level. MPTCP endpoints deallocate all resources (the token mapping and \textit{MpTcpSocketBase} object are also deleted from the \textit{TcpL4Protocol} object) only when all subflows have been closed.
\\
\section{Current State and Future Work}
Our MPTCP model is in a fully working state. A few minor additions are still required to fully comply with the RFC 6824 (e.g. \textsc{mp-capable} does not exchange keys for securing a connection and \textsc{mp-prio} is not implemented). To showcase our model we present results from some simple simulation scenarios. Figure \ref{fig:sub1} illustrates the evolution of the congestion window when an MPTCP connection with a single subflow is opened between two nodes connected via a point-to-point link. For clarity, we haven't plotted the results from simulations using the standard TCP NewReno model; they were identical to the ones presented in the figure. In Figure \ref{fig:sub2} we plot the congestion window progression, when two subflows are used via two point-to-point links. We have also examined the behaviour of our NewReno loss recovery mechanism by reconstructing a simulation scenario from \cite{Fall_comparisons}. Our results in \ref{fig:sub3} are very similar to Figure 3 in \cite{Fall_comparisons}.

This work is part of our larger effort towards building an ns-3 based simulation platform for experimenting with transport protocols and congestion control algorithms in data centres. We plan to make necessary changes in the core ns-3 code so that congestion control and loss recovery mechanisms can be applied to different transport protocols by reusing existing code. For example we had to rewrite the NewReno loss recovery code in our \textit{MpTcpSocketBase} class, although we could have reused code from the \textit{TcpNewReno} class. A major challenge towards building a simulation platform for data centres is the scale of simulations. We are currently working on identifying the current limits and the required changes to scale up to large simulated topologies.

\bibliographystyle{abbrv}
\bibliography{mptcp-wns3-2014}

\balancecolumns

\end{document}